\documentclass[pss]{wiley2sp}
\usepackage{amsmath}

\begin{document}

\title{Current Density Functional Theory for one-dimensional fermions}

\titlerunning{CDFT for one-dimensional fermions}

\author{%
  Michael Dzierzawa,
  Ulrich Eckern,
  Stefan Schenk\textsuperscript{\textsf{\Ast}},
  and Peter Schwab}

\authorrunning{M.\ Dzierzawa et al.}

\mail{e-mail
  \textsf{stefan.schenk@physik.uni-augsburg.de}, phone
  +49 821 598-3240, fax +49 821 598-3262}

\institute{%
  Institut f\"ur Physik, Universit\"at Augsburg, 86135 Augsburg, Germany}



\abstract{%
%
%
%
\abstcol{%
The frequency-dependent response of a one-dimensional fermion system is 
investigated using Current Density Functional Theory (CDFT) 
within the local approximation (LDA). DFT-LDA, and in particular
CDFT-LDA, reproduces very well the dispersion of the collective excitations.
Unsurprisingly, however, the approximation fails for details of the dynamic 
response for large wavevectors.
}{%
In particular, we introduce CDFT for the one-dimensional spinless fermion
model with nearest-neighbor interaction,
and use CDFT-LDA plus exact (Bethe ansatz) results for the 
groundstate energy as function of particle density and boundary phase to
determine the linear response. The successes and failures of this
approach are discussed in detail.
}}

%
%

\maketitle   

\section{Introduction}

Density Functional Theory (DFT) is an efficient and powerful tool for
determining the electronic structure of solids.
While originally developed for continuum electron systems with Coulomb
interaction \cite{hohenberg1964,kohn1965},
DFT has also been applied to lattice models, such as the Hubbard
model \cite{gunnarsson1986,schonhammer1987,schonhammer1995,lima2003},
in order to develop new approaches to correlated electron systems:
lattice models often allow for exact solutions which hence can serve as
benchmarks for assessing the quality of approximations.

Very useful for applications is the Local Density Approximation (LDA)
where the exchange-correlation energy of the inhomogeneous system
under consideration is constructed via a local approximation from the
homogeneous electron system. A lattice version of LDA has been 
suggested for one-dimensional systems \cite{schonhammer1995} where the 
underlying homogeneous system can be solved using Bethe ansatz. For
recent applications of Bethe ansatz LDA, see also, for example, Refs.\
\cite{lima2003,lima2002,xianlong2006,xianlong2007,alcaraz2007,verdozzi2007,li2008}.

In addition, the time-dependent version of DFT has been developed and
applied \cite{runge1984,gross1985}, in particular, the current density
version \cite{vignale1996}; a recent review \cite{onida2002} and book 
\cite{marques2006} provide excellent overviews, including the relation
to standard many-body Green's function approaches.

In this article, we focus on the one-dimensional spinless fermion model
with nearest-neighbor interaction, which is exactly solvable in the
homogeneous case \cite{yang1966,haldane1980}. We extend our recent
DFT-LDA approach \cite{schenk2008} to 
{\em current} density functional theory \cite{vignale1996}. We
present the model in Sect.\ 2, and discuss general aspects of linear
response theory in Sect.\ 3. CDFT and LDA are presented in Sect.\ 4; for
simplicity we use the static zero-temperature limit in order to present the
general ideas. Our results are given in Sect.\ 5, and brief conclusions in 
Sect.\ 6.

In the following, $\hbar$ as well as the lattice constant are put equal
to one; the system size is denoted by $L$, and we assume periodic boundary
conditions.

\section{The model}

We consider one-dimensional spinless fermions described by the Hamiltonian
\begin{equation} 
\label{eq1}
\hat H = \hat T + \hat V + \sum_l v_l \hat n_l
\end{equation}
where
\begin{equation} 
\label{eq2}
\hat T = - t \sum_l 
\left( {\rm e}^{i\phi_l} \hat c_l^+ \hat c_{l+1} + {\rm h.c.} \right)
\end{equation}
is the kinetic energy, and
\begin{equation}
\label{eq3}
\hat V = V \sum_l \hat n_l \hat n_{l+1}
\end{equation}
the interaction. Generally, the phases $\{\phi_l\}$ and local potentials
$\{ v_l \}$ can be time-dependent. The hat denotes operator-valued
quantities. Clearly
\begin{equation}
\label{eq4}
\hat n_l = \partial \hat H / \partial v_l \; , \;
\hat \jmath_l = \partial \hat H / \partial \phi_l \; ,
\end{equation}
where
\begin{equation}
\label{eq5}
\hat \jmath_l = -it
\left( {\rm e}^{i\phi_l} \hat c_l^+ \hat c_{l+1} - {\rm h.c.} \right)
\end{equation}
denotes the local current. Particle conservation follows immediately by
noting that in the Heisenberg picture, we have
\begin{equation}
\label{eq6}
\dot{\hat n}_l = i [{\hat H},{\hat n}_l] 
= - (\hat \jmath_l - \hat \jmath_{l-1}) \; .
\end{equation}
A gauge transformation is described by the (unitary) operator
\begin{equation}
\label{eq7}
\hat U = \exp (i \sum_l \chi_l \hat n_l )
\end{equation}
such that the time development of the transformed wavefunctions is 
determined by
\begin{equation}
\label{eq8}
{\hat {\tilde H}} = {\hat U} {\hat H} {\hat U}^+
                  - \sum_l {\dot\chi}_l \hat n_l \; .
\end{equation}
Thus a gauge transformation implies the replacements
$\phi_l \to \phi_l + \chi_l - \chi_{l+1}$ and
$v_l \to v_l - {\dot\chi}_l$. Note that the combinations
$e_l = {\dot\phi}_l - (v_{l+1} - v_l)$, corresponding to the
electric field in electrodynamics, as well as $\Phi = \sum_l \phi_l$,
corresponding to the magnetic flux \cite{eckern1995}, are gauge invariant.

In the context of DFT, we will also introduce an auxiliary single-particle
system ``s'', defined by
\begin{equation}
\label{eq9}
\hat H^{\rm s} = \hat T^{\rm s} + \sum_l v_l^{\rm s} \hat n_l
\end{equation}
where $T^{\rm s}$ is obtained from $T$ defined in (\ref{eq2}) by the
replacement $\phi_l \to \phi_l^{\rm s}$ for all $l$.

\section{Linear response}

Using standard techniques \cite{kubo1991} we discuss briefly the 
linear response to
time-dependent perturbations $\{ \delta v_l(t) \}$, $\{ \delta \phi_l(t) \}$
such that
\begin{equation}
\label{eq10}
\delta \hat H = \sum_l (\hat n_l \delta v_l + \hat \jmath_l \delta \phi_l) \; .
\end{equation}
The response function $\check\chi$, defined through the relation
\begin{equation}
\label{eq11}
\left( \begin{array}{c} \delta\langle \hat n_l \rangle (t) \\
                        \delta\langle \hat \jmath_l \rangle (t)
       \end{array} \right)
= - \sum_m \int_{-\infty}^{+\infty} \! dt^\prime
{\check\chi}^{lm} (t-t^\prime)
\left( \begin{array}{c} \delta v_m (t^\prime) \\
                        \delta \phi_m (t^\prime)
       \end{array} \right) \; ,
\end{equation}
has four entries, $\chi_{nn}$, $\chi_{nj}$, $\chi_{jn}$, and $\chi_{jj}$
which each are $L\times L$ matrices with respect to the lattice sites
$\{lm\}$; these are related to expectation values of commutators,
for example
\begin{equation}
\label{eq12}
\chi_{nn}^{lm} (t-t^\prime) =
i \Theta (t-t^\prime) \langle [\hat n_l (t), \hat n_m (t^\prime) ]\rangle \; 
\end{equation}
where $\Theta (t-t^\prime)$ is the unit step function, and $\langle\dots\rangle$
denotes the quantum statistical average. Considering a Fourier
transformation with respect to the time difference, $t-t^\prime \to \omega$,
the response functions obey Onsager's relations:
\begin{equation}
\label{eq13}
\chi_{nn}^{lm} (\omega; \{\phi_l\}) 
= \chi_{nn}^{ml} (\omega; \{-\phi_l\}) 
\end{equation}
\begin{equation}
\label{eq14}
\chi_{nj}^{lm} (\omega; \{\phi_l\}) 
= - \chi_{jn}^{ml} (\omega; \{-\phi_l\}) 
\end{equation}
\begin{equation}
\label{eq15}
\chi_{jj}^{lm} (\omega; \{\phi_l\}) 
= \chi_{jj}^{ml} (\omega; \{-\phi_l\}) 
\end{equation}
The minus sign in (\ref{eq14}) reflects that the current is odd under
time reversal. The density-density, density-current and current-current
response functions are related to each other due to particle conservation.

Considering a homogeneous situation, i.e., 
$\phi_l = \phi$ and $v_l = 0$ for all $l$, we obtain
\begin{equation}
\label{eq16}
\chi_{nn} (q,\omega; \phi) 
= \chi_{nn} (-q,\omega; -\phi)
\end{equation}
\begin{equation}
\label{eq17}
\chi_{nj} (q,\omega; \phi) 
= - \chi_{jn} (-q,\omega; -\phi)
\end{equation}
\begin{equation}
\label{eq18}
\chi_{jj} (q,\omega; \phi) 
= \chi_{jj} (-q,\omega; -\phi)
\end{equation}
where $q$ is the wavevector.
For the homogeneous single-particle system (\ref{eq9}) the explicit results
are as follows:
\begin{equation}
\label{eq19}
\chi^{\rm s}_{\alpha\beta} (q,\omega; \phi) =
- \frac{1}{L} \sum_k 
\frac{n_k - n_{k+q}}{\omega + \epsilon_k - \epsilon_{k+q} + i0}
\, \kappa_{\alpha\beta}
\end{equation}
where $\kappa_{nn} = 1$, $\kappa_{nj} = \kappa_{jn} = v_{k+q/2}$, 
and $\kappa_{jj} = v^2_{k+q/2}$. In addition, 
$\epsilon_k = -2t \cos (k+\phi)$ is the
free-particle dispersion, $v_k = \partial\epsilon_k /\partial k$ the 
corresponding velocity, and $n_k$ denotes the Fermi function. 
We note also that 
\begin{equation}
\label{eq20}
\chi_{nj} (q,\omega; \phi) = \chi_{jn} (q,\omega; \phi)
\end{equation}
due to parity symmetry. Particle conservation implies
\begin{equation}
\label{eq21}
\omega \cdot \chi_{nn} = 2 \sin (q/2) \cdot \chi_{nj}
\end{equation}
and
\begin{equation}
\label{eq22}
\omega^2 \cdot \chi_{nn} = [2 \sin (q/2)]^2 \cdot \chi_{jj}
\end{equation}
where we suppressed the arguments ($q,\omega;\phi$) for simplicity. These 
relations allow writing the density and current response in gauge-invariant
form, for example:
\begin{equation}
\label{eq23}
\delta n (q,\omega) =
\frac{\chi_{nn} (q,\omega)}{2i \sin (q/2)} \cdot e (q,\omega)
\end{equation}
where $e (q,\omega) = -i\omega \phi (q,\omega) - 2i \sin (q/2) v (q,\omega)$;
compare the discussion below (\ref{eq8}). Equivalently
\begin{equation}
\label{eq24}
\delta j (q,\omega) =
\frac{\chi_{jj} (q,\omega)}{i\omega} \cdot e (q,\omega) \;
\end{equation}
such that $\omega\cdot\delta n(q,\omega) = 2\sin(q/2) \cdot\delta j (q,\omega)$.

\section{Current Density Functional Theory}

In this section, we briefly outline the current density functional theory and
the local density approximation, without discussing questions of uniqueness
and other mathematical difficulties \cite{levy1982}. For simplicity of notation,
we restrict ourselves to the static zero-temperature limit. The generalization
to the time-dependent finite-temperature case is straightforward, utilizing
generating functionals, functional derivatives, etc \cite{fukuda1994}. 

\subsection{General aspects}

We start with
the groundstate energy of the Hamiltonian (\ref{eq1}), $E$,
which is a function of the local phases $\{\phi_l\}$ and potentials $\{v_l\}$
with the property
\begin{equation}
\label{eq25}
n_l = \langle \hat n_l \rangle = \partial E / \partial v_l \; , \;
j_l = \langle \hat \jmath_l \rangle = \partial E / \partial \phi_l \; .
\end{equation}
We transform to new variables $\{n_l\}$ and $\{j_l\}$, i.e., introduce the
Legendre transform, $F$, according to the relation
\begin{equation}
\label{eq26}
F = E - \sum_l ( v_l n_l + \phi_l j_l )
\end{equation}
such that
\begin{equation}
\label{eq27}
v_l = - \partial F / \partial n_l \; , \; 
\phi_l = - \partial F / \partial j_l \; .
\end{equation}
Obviously, some care will be necessary due to gauge invariance. 
For the static DFT case, this is a minor problem: 
a constant can be added to the local potentials without essential changes 
of the physics. In the general
dynamic case, one has to keep in mind that density and current are not
independent, but related by particle conservation.

In the next step, we perform an analogous Legendre transformation for the
auxiliary single-particle system (\ref{eq9}),
\begin{equation}
\label{eq28}
F^{\rm s} = 
E^{\rm s} - \sum_l ( v^{\rm s}_l n_l + \phi^{\rm s}_l j_l ) \; .
\end{equation}
Then, by definition, we have
\begin{equation}
\label{eq29}
v^{\rm s}_l = v_l + \frac{\partial E^{\rm HXC}}{\partial n_l} \; , \;
\phi^{\rm s}_l = \phi_l + \frac{\partial E^{\rm HXC}}{\partial j_l}
\end{equation}
where $E^{\rm HXC} \equiv F - F^{\rm s}$. (The superscript ``HXC'' refers to 
Hartree-Exchange-Correlation.) Introducing explicitly the Hartree
contribution, $E^{\rm H}$, through
\begin{equation}
\label{eq30}
E^{\rm HXC} = E^{\rm H} + E^{\rm XC}
\end{equation}
with $E^{\rm H} = V \sum_l n_l n_{l+1}$, we arrive at the standard relation
\begin{equation}
\label{eq31}
v^{\rm s}_l = v_l + v^{\rm H}_l + v^{\rm xc}_l
\end{equation}
where presently $v^{\rm H}_l = V(n_{l+1} + n_{l-1})$; the
exchange-correlation potential is given by
$v^{\rm xc}_l = \partial E^{\rm XC} / \partial n_l$. In addition,
\begin{equation}
\label{eq32}
\phi^{\rm s}_l = \phi_l + \phi^{\rm xc}_l \; , \;
\phi^{\rm xc}_l = \frac{\partial E^{\rm XC}}{\partial j_l} \; ;
\end{equation}
here only $E^{\rm XC}$ appears
since the Hartree energy does not depend on the currents. Explicitly:
\begin{equation}
\label{eq33}
E^{\rm XC} = \langle 0 | \hat T + \hat V | 0 \rangle
            -\langle 0_{\rm s} | \hat T^{\rm s} | 0_{\rm s} \rangle
	    - E^{\rm H} + \sum_l \phi^{\rm xc}_l j_l 
\end{equation}
where $|0\rangle$ and $|0_{\rm s}\rangle$ are the groundstate wavefunctions of
the interacting and the single-particle system, respectively.

\subsection{Local approximation}

In the next step, an approximation for $E^{\rm XC}$ is needed. As usual, a local
approximation is employed, according to the following recipe:
(i) consider the static, homogeneous case, and determine $E^{\rm XC} (n,j)$;
(ii) define $\epsilon^{\rm xc} = E^{\rm XC} / L$; (iii) approximate 
$E^{\rm XC} (\{n_l,j_l\})$ by $\sum_l \epsilon^{\rm xc} (n_l ,j_l )$. 

For step
(i), we note that the first term in (\ref{eq33}), henceforth denoted 
$E^{\rm BA}$, is known from the Bethe ansatz
solution \cite{yang1966,haldane1980} of the model, albeit
as a function of $n$ and $\phi$. Thus the phase variable has to be eliminated
from this expression in favor of the current, using the relation $j =
\partial E^{\rm BA} (n,\phi) /\partial \Phi$ (recall that $\Phi = L\phi$). In
relation to the second term in (\ref{eq33}), which we denote $E^0$, we recall
the single-particle result ($-\pi/L < \phi < \pi/L$)
\begin{equation}
\label{eq34}
E^0 (\phi) = E^0 (0) \cdot \cos \phi \; , \;
E^0 (0) \simeq - \frac{2t}{\pi} L \sin (\pi n)
\end{equation}
where the latter relation holds for large $L$. Since $\phi\sim 1/L$, we may
expand for small $\phi$; in particular, the Drude weights, $D^{\rm BA}$ and
$D^0$, are defined according to the following relations ($\Phi\to 0$):
\begin{equation}
\label{eq35}
E^{\rm BA} (\Phi) - E^{\rm BA} (0) = D^{\rm BA} \Phi^2 / L 
\end{equation}
\begin{equation}
\label{eq36}
E^0 (\Phi) - E^0 (0) = D^0 \Phi^2 / L
\end{equation}
Note that $D^{\rm BA}$ and $D^0$ are functions of the density, and 
$D^{\rm BA}$ depends on the interaction $V$. For example,
$D^0 = (t/\pi) \sin (n\pi) = v_F /2\pi$, where $v_F$ is the bare Fermi
velocity, and $D^{\rm BA} = \pi t \sin \mu /[4\mu (\pi -\mu)]$ for half
filling ($n=1/2$), where $V$ (in the range $-2t \ldots 2t$) is related to
$\mu$ by $V = -2t \cos \mu$.

Combining the above relations, we obtain
\begin{equation}
\label{eq37}
E^{\rm XC} (n,j) = E^{\rm BA} (n,0) - E^0 (n,0) - L Vn^2
                 + \frac{L}{2} \lambda^{\rm xc} j^2
\end{equation}
where
\begin{equation}
\label{eq38}
\lambda^{\rm xc} (n) = \frac{1}{2} \left( \frac{1}{D^0 (n)} - \frac{1}{D^{\rm BA} (n)} \right) \; .
\end{equation}
Note that $\lambda^{\rm xc} (n) \le 0$ since $D^{\rm BA} (n) \le D^0 (n)$. The next
steps, (ii) and (iii), are straightforward. The resulting approximation may be
called CDFT-LDA.

\subsection{CDFT-LDA and linear response}

For the determination of the response functions, we again employ the auxiliary
single-particle system as follows. First, we consider the response of the 
s-system to small variations $\delta v^{\rm s}$, $\delta \phi^{\rm s}$, thereby
defining the quantity ${\check\chi}^{\rm s}$ analogous to (\ref{eq11}).
(Here and in
the following two equations, we will resort to a short-hand notation.)
Second, we take the variations of the Hartree and the
exchange contributions to the potentials and phases into account, according
to
\begin{equation}
\label{eq39}
\left( \begin{array}{c} \delta v^{\rm s} \\
                        \delta \phi^{\rm s} 
       \end{array} \right)
=        
\left( \begin{array}{c} \delta v \\
                        \delta \phi 
       \end{array} \right)
+
{\check{\rm f}}^{\rm HXC}
\left( \begin{array}{c} \delta n \\
                        \delta j 
       \end{array} \right)
\end{equation}
thereby introducing ${\check{\rm f}}^{\rm HXC} 
= {\check{\rm f}}^{\rm H} + {\check{\rm f}}^{\rm XC}$. The result is
\begin{equation}
\label{eq40}
{\check\chi} = \left(\check 1 + \check\chi^{\rm s} {\check{\rm f}}^{\rm HXC}\right)^{-1}
              \check\chi^{\rm s} \; ,
\end{equation}	      
which reduces to the standard RPA expression when ${\check{\rm f}}^{\rm XC} = 0$. 
Note that ${\check{\rm f}}^{\rm H}$ only has an $nn$-entry, in Fourier
representation given by $V(q) = 2V \cos q$.

\begin{figure}[t]%
\includegraphics*[width=\linewidth]{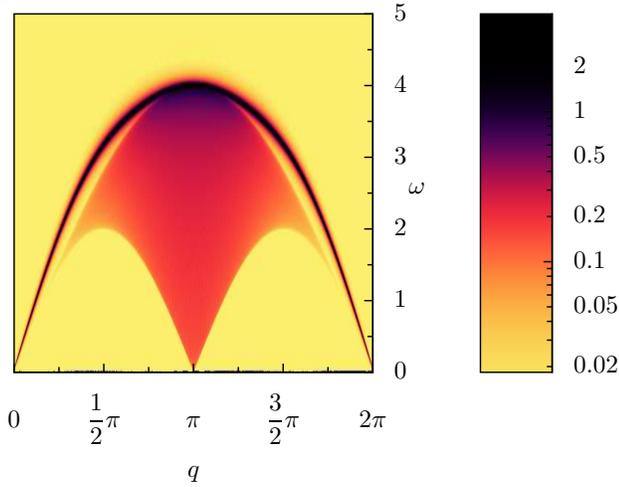}
\caption{%
Contour plot of the imaginary part of the dynamical susceptibility
${\rm Im}\, \chi_{nn} (q,\omega)$ (in units of $t^{-1}$), as obtained from
(\ref{eq41}), for half-filling and $V/t = 1$. Here and in the 
following figures, $\omega$ is given in units of $t$. For the plot, we replace
the imaginary part of the denominator in (\ref{eq19}) by a finite value, $\eta$,
which we choose here to be $0.005t$.}
\label{fig1}
\end{figure}

The above expression is exact provided the exact functional 
${\check{\rm f}}^{\rm XC}$ 
is used. In the following, however, we rely on the results of the
previous subsection, and consider in particular $j \to 0$; in this
limit, see above, we may use ${\rm f}^{\rm HXC}_{nj} \simeq 0$, 
${\rm f}^{\rm HXC}_{jn} \simeq 0$. Employing particle conservation again, 
we find
\begin{equation}
\label{eq41}
\chi_{nn} (q,\omega) = 
\frac{\chi^{\rm s}_{nn}}{1 + \chi^{\rm s}_{nn} \left( {\rm f}^{\rm HXC}_{nn} + \frac{\omega^2}{4\sin^2 (q/2)} {\rm f}^{\rm HXC}_{jj} \right)}
\end{equation}
where $\chi^{\rm s}_{nn}$, of course, depends on $q$ and $\omega$.
If in addition ${\rm f}^{\rm HXC}_{jj} = 0$, we recover
the approximation known as adiabatic LDA \cite{schenk2008,remark1}.

In order to discuss the result (\ref{eq41}) in more detail, recall that in
the long-wavelength low-frequency limit the density response of the s-system
is given by
\begin{equation}
\label{eq42}
\chi^{\rm s}_{nn} (q,\omega) \simeq \chi^{\rm s}_{\rm stat}
  \frac{(qv_F)^2}{(qv_F)^2 - (\omega +i0)^2}
\end{equation}
where the static susceptibility $\chi^{\rm s}_{\rm stat} = 1 / \pi v_F$. 
Considering the limit $\omega = 0$, $q \to 0$, and noting that
\begin{equation}
\label{eq43}
{\rm f}^{\rm XC}_{nn} = \frac{1}{L} \left( \frac{\partial^2 E^{\rm BA}}{\partial n^2} - \frac{\partial^2 E^0}{\partial n^2} - 2 L V \right)
\end{equation}
it is straightforward to verify that
\begin{equation}
\label{eq44}
\chi_{\rm stat} = \left( \frac{1}{L} \frac{\partial^2 E^{\rm BA}}{\partial n^2} \right)^{-1} \; .
\end{equation}
On the other hand, taking $q \to 0$ first, we find
\begin{equation}
\label{eq45}
\chi_{jj} (q=0, \omega \to 0) = - 2 D^{\rm BA} \; ,
\end{equation}
i.e., the exact result. The minus sign here is due to our definition of the
response function, compare (\ref{eq11}). Inserting (\ref{eq42}) into
(\ref{eq41}), we find (for small $q$, $\omega$)
\begin{equation}
\label{eq46}
\chi_{nn} (q,\omega) \simeq \chi_{\rm stat}
  \frac{(qv)^2}{(qv)^2 - (\omega +i0)^2}
\end{equation}
where
\begin{equation}
\label{eq47}
v^2 = v^2_F \frac{1 + {\rm f}^{\rm HXC}_{nn} /\pi v_F}{1 - {\rm f}^{\rm HXC}_{jj} v_F / \pi} = \frac{2 D^{\rm BA}}{\chi_{\rm stat}}
\end{equation}
which -- as to be expected in view of (\ref{eq45}) -- is the exact expression.

The numerical results presented below are based on Eqs.\ (\ref{eq37}) and
(\ref{eq41}).

\section{Numerical results}

In Fig.\ \ref{fig1} we show a contour plot of the imaginary part of $\chi_{nn}
(q,\omega)$ for half-filling and $V/t =1$. The apparent continuum of 
excitations can be identified with the particle-hole continuum; its spectral 
weight vanishes in the long-wavelength limit. Above the continuum, 
there is a well-defined branch of collective excitations; as discussed above, the
corresponding velocity for $q \to 0$ has the exact value. However, the contour
plot is almost indistinguishable from the corresponding one obtained within
adiabatic LDA (compare Fig.\ 4 in \cite{schenk2008}). 

\begin{figure}[t]%
\includegraphics*[width=\linewidth]{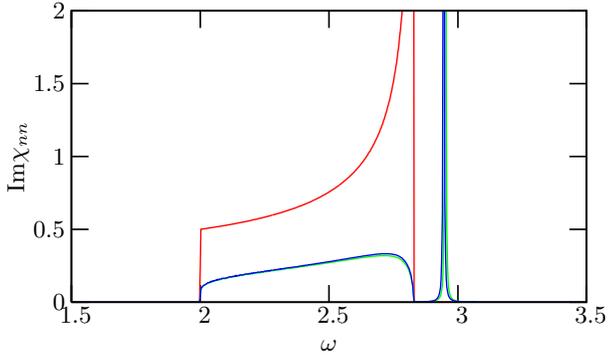}
\caption{%
${\rm Im}\, \chi_{nn} (\pi/2,\omega)$ versus $\omega$ for half-filling,
comparing DFT-LDA (green) with CDFT-LDA (blue) for $V = 0.5 t$. The
non-interacting case, $V = 0$ (red), is given as reference. ($\eta = 0.0001t$)}
\label{fig2}
\end{figure}

\begin{figure}[t]%
\includegraphics*[width=\linewidth]{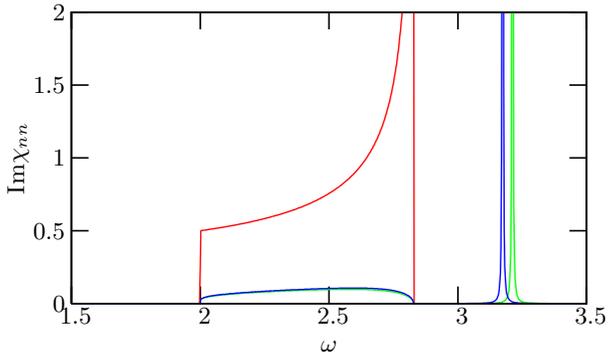}
\caption{%
Same as Fig.\ \ref{fig2}, for $V = t$.}
\label{fig3}
\end{figure}

Thus, in order to highlight the differences between adiabatic LDA \cite{schenk2008}
and the present CDFT-LDA, we present in Figs.\ \ref{fig2} and \ref{fig3}
${\rm Im}\, \chi_{nn} (q,\omega)$ for a fixed wavevector, $q = \pi/2$, as a
function of frequency, again for half-filling. In Fig.\ \ref{fig2} ($V/t = 0.5$),
even though the interaction is still moderate, the spectral weight of the continuum
is already strongly reduced. However, the frequency range $\omega^0_{-} \ldots
\omega^0_{+}$, where $\omega^0_{-} = 2t|\sin q |$ and $\omega^0_{+} = 4t\sin (q/2)$,
is fixed and equals the range of the non-interacting model for all $V$. Above 
the continuum, the collective mode is apparent; for this interaction parameter,
however, the correction due to CDFT-LDA is minimal.

With increasing interaction, the spectral weight of the continuum is further reduced,
and the frequency of the collective mode is shifted slightly to a lower value
compared to adiabatic LDA, consistent with (\ref{eq47}); see Fig.\ \ref{fig3}.

The above results have to be contrasted with recent exact results for the dynamic
response of the spinless fermion model 
\cite{pereira2006,pustilnik2006,pereira2008,cheianov2008}, which -- 
unsurprisingly -- are not correctly
reproduced within CDFT-LDA. For example, a continuum of collective excitations is
found in \cite{pustilnik2006,pereira2008} for a certain frequency range,
$\omega_{-} < \omega < \omega_{+}$,
where $\omega_\pm$ are $q$- and {\em interaction}-dependent. In addition, spectral 
weight is shifted (for a repulsive interaction) to the lower end of the continuum, 
leading to a power-law divergence near $\omega_{-}$. For an attractive interaction, 
on the other hand, a bound state is found in the dynamical structure factor
above the continuum \cite{pereira2008,remark2}.

\begin{figure}[htb]%
\includegraphics*[width=\linewidth]{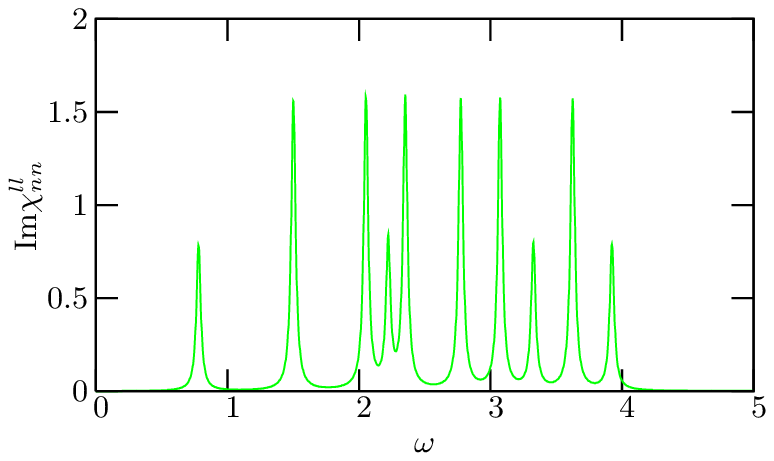}

a) $V = 0$ \bigskip

\includegraphics*[width=\linewidth]{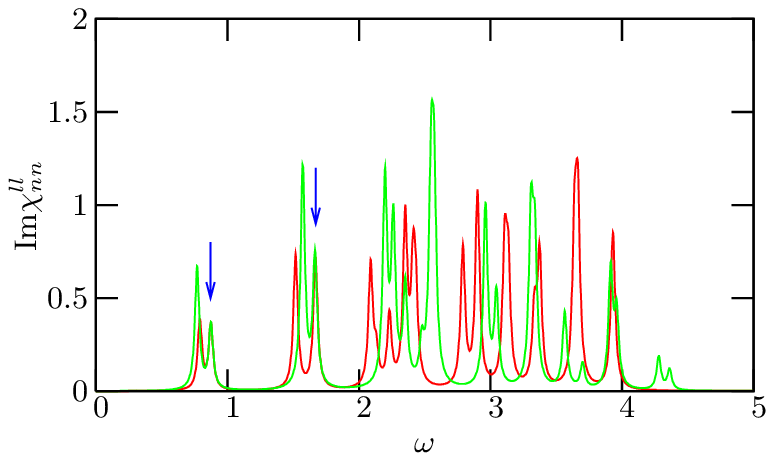}

b) $V = 0.4 t$ \bigskip

\includegraphics*[width=\linewidth]{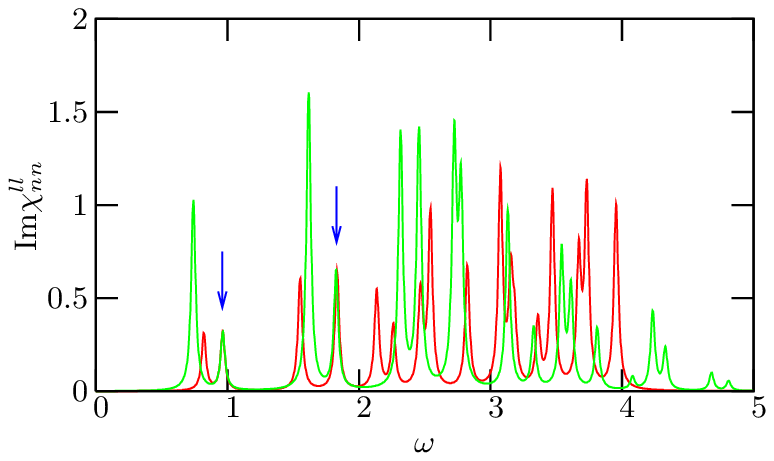}

c) $V = 0.8 t$

\caption{%
Local density response ${\rm Im}\, \chi^{ll}_{nn} (\omega)$
versus $\omega$ for a small ($L = 16$)
system (at half-filling), comparing exact diagonalization results (green)
with CDFT-LDA (red). Here $\eta = 0.02$. The blue arrows indicate excitations
that are obtained almost exactly within CDFT-LDA, i.e., near these peaks the
green and the red curves are on top of each other on the scale of this plot.}
\label{fig4}
\end{figure}

In order to clarify how the differences between exact and CDFT-LDA results develop
with increasing interaction, we have performed exact diagonalization studies for
small systems of 16 lattice sites. Some of our results are shown in Fig.\ \ref{fig4},
where we plot the imaginary part of the local density response function
\begin{equation}
\label{eq48}
{\rm Im}\, \chi^{ll}_{nn} (\omega) = L^{-1} \sum_q {\rm Im}\, \chi_{nn} (q,\omega)
\end{equation}
for half-filling versus frequency, for $V = 0$, $ 0.4t$, and $0.8t$. 

The usefulness of ${\rm Im}\, \chi_{nn}^{ll}(\omega)$ lies in the fact that 
the excitation energies of the interacting system are given by the
poles of the response function, which appear as broadened 
$\delta$-peaks in the figures due to the finite value $\eta = 0.02t$ of 
the imaginary part of the frequency.
Figure 4a shows the susceptibility of the non-interacting system, as reference. 
Obviously exact diagonalization and CDFT-LDA yield identical results in this case.

In Fig.\ 4b, where $V/t = 0.4$, the two peaks with lowest energy are split
into doublets. Remarkably, for each doublet the position
of the peak with the higher energy (marked by blue arrows)
is almost exactly obtained within CDFT-LDA. 
These peaks correspond to the two lowest possible $q$-values for a 
16 site system, $\pi/8$ and $\pi/4$, respectively,
which again demonstrates the validity of the CDFT-LDA in the
long-wavelength limit.

On the other hand, the peaks with lower energy within the doublets,
corresponding to $q$-values near $\pi$, are clearly off.
A further feature that is not obtained within CDFT-LDA is the appearance
of high energy satellites beyond the upper limit $\omega_+^{0} = 4t$
of the non-interacting continuum. 

The trends already apparent for $V/t = 0.4$ become even clearer for
$V/t = 0.8$ (Fig.\ 4c). In addition, in the exact data more and more 
spectral weight is shifted down to the left sub-peaks of the low energy 
doublets, a feature which is not obtained within CDFT-LDA.
The transfer of spectral weight to lower frequencies
eventually leads to the formation
of the power-law divergence at the lower end of the continuum in
the infinite system \cite{pustilnik2006,pereira2008}.

\section{Conclusion and outlook}

We have demonstrated that Bethe ansatz current density functional
theory correctly describes the Luttinger liquid properties of the 
one-dimen\-si\-onal
spinless fermion model in the long-wavelength low-frequency limit, in particular,
both limits -- $\omega = 0, q \to 0$ and $\omega \to 0, q = 0$ -- are recovered;
compare (\ref{eq44}) and (\ref{eq47}). The local approximation for the
exchange-correlation potential is insufficient for other aspects: For the static
case, it misses the ``critical'' properties related to $2k_F$-scattering
\cite{schenk2008}, while for the dynamic response, 
the excitation spectrum for large wavevectors is not correct.
It is unclear at this moment, at least to us, whether
some of the shortcomings of CDFT-LDA can be cured, for example, by pursuing the
so-called exact-exchange potential approach \cite{langreth1980}.

The static case has been discussed in our recent work \cite{schenk2008} in
considerable detail. We already noted that the $q \to 0$ limit of the static 
response is, by construction, obtained exactly within LDA; compare the
discussion in connection with (\ref{eq44}). Furthermore, as
a major improvement in comparison with the Hartree approximation, Bethe
ansatz LDA correctly predicts a non-charge-ordered groundstate for a large
range of parameters. The static density response was found to agree very well
with the exact result for not too large systems, low particle density, and
wavevectors $q < 2k_F$.

Good agreement between exact (density matrix renormalization group) 
calculations and time-dependent DFT is also reported
in a recent study \cite{li2008} of the collective density and spin dynamics
of the one-dimensional Hubbard model. In this work,
the adiabatic local spin-density
approximation is employed to investigate the density and spin response to
a local time-dependent perturbation for small and relatively dilute
systems. 

On the other hand, it is obvious that the exact exchange-correlation 
potential can be obtained by numerical methods, at 
least for relatively small model systems. In this context,
we noted previously \cite{schenk2008} that the Bethe ansatz LDA combined with
numerically determined exact exchange-correlation potentials 
\cite{schmitteckert2008} might be a useful approach for short, inhomogeneous 
systems, like quantum dots and mo\-le\-cules. Ab initio calculations of the 
linear conductance through molecules, e.g., as a function
of the gate voltage, are usually based on DFT-LDA -- but the theoretical and
experimental results differ typically by an order of magnitude, which seems to
be related to the quality of the xc-potentials employed; see \cite{toher2005}
for detailed discussions.

Following this suggestion \cite{schmitteckert2008}, we have started an 
investigation of small one-dimensional interacting dots, typically consisting 
of five sites, coupled to one-dimensional leads, within (i) LDA, (ii) 
Hartree-Fock approximation, and (iii) DFT with the exchange-correlation 
potential being determined from exact diagonalization.
Preliminary results indicate that the 
``exact DFT'' leads to a considerable improvement compared to LDA and 
Hartree-Fock; i.e., good agreement is obtained with the density matrix 
renormalization group studies of \cite{schmitteckert2008}. A realistic 
calculation of transport properties hence
seems to be feasible, by combining ``exact DFT'' for small dots with LDA for the
leads, which do not necessarily need to be one-dimensional in this approach.

\begin{acknowledgement}
This work was supported by the Deut\-sche Forschungsgemeinschaft 
through SFB 484.
\end{acknowledgement}

\end{document}